# *In Situ* Compensation Method for Precise Integral SQUID Magnetometry of Miniscule Biological, Chemical, and Powder Specimens Requiring the Use of Capsules

**Katarzyna Gas and Maciej Sawicki**

Institute of Physics, Polish Academy of Sciences, Aleja Lotnikow 32/46, PL-02668 Warsaw, Poland
kgas@ifpan.edu.pl (K.G.); mikes@ifpan.edu.pl (M.S.)

**Abstract:** Steadily growing interest in magnetic characterization of organic compounds for therapeutic purposes or of other irregularly shaped specimens calls for refinements of experimental methodology to satisfy experimental challenges. Encapsulation in capsules remains the method of choice, but its applicability in precise magnetometry is limited. This is particularly true for minute specimens in the single milligram range as they are outweighed by the capsules and are subject to large alignment errors. We present here a completely new experimental methodology that permits 30-fold *in situ* reduction of the signal of capsules by substantially restoring the symmetry of the sample holder that is otherwise broken by the presence of the capsule. In practical terms it means that the standard 30 mg capsule is seen by the magnetometer as approximately a 1 mg object, effectively opening the window for precise magnetometry of single milligram specimens. The method is shown to work down to 1.8 K and in the whole range of the magnetic fields. The method is demonstrated and validated using the reciprocal space option of MPMS-SQUID magnetometers; however, it can be easily incorporated in any magnetometer that can accommodate straw sample holders (i.e., the VSM-SQUID). Importantly, the improved sensitivity is accomplished relying only on the standard accessories and data reduction method provided by the SQUID manufacturer, eliminating the need for elaborate raw data manipulations.

---

## 1. Introduction

Commercial general-purpose superconducting quantum interference devices (SQUID) have become ubiquitous experimental tools in laboratories where magnetic characterization and investigations of nanometer-scale objects is important and frequently indispensable in providing the vital insight into investigations. The range of objects include, in particular, ultrathin films [1,2], nanoparticles [3,4], and inclusions in magnetically dilute systems [5–7], 2D [8,9], and 1D [10,11] systems, or systems in which their surface matters [12], like topological insulators [13–15]. In the prevailing number of cases, the materials of interest are on a solid state substrate, which make the mounting of the sample relatively easy; a breadth of relevant experimental codes has been elaborated and methods for eliminating various pitfalls have been presented [16–20], which were reviewed by Pereira [21]. In this context, a new window of experimental opportunities was opened by the elaboration of methodology and full quantification of the *in situ* substrate compensational method [22], which allows for truly precise integral magnetometry of nanometer scale objects with absolute sensitivity down to $10^{-8}$ emu [23] or of thin antiferromagnetic layers [24].

However, there exists a wide spectrum of subjects in which SQUID magnetometry cannot be performed according to the rules elaborated for solid state objects. These are shapeless bodies, like irregular small bits or fragments, powdered materials, generally understood soft condensed materials [25], and all other chemical compounds and biological specimens, frequently of a strong spintronics [26] or therapeutic relevance [27–30]. These systems usually exhibit very weak magnetic properties, and they are predominantly diamagnets. Only a handful of them allow for the use of methods developed within the solid state approach to magnetometry—these that can be bound by weakly magnetic glues without any degradation effects. One of such very con-venient bonding agents is the so-called GE varnish [31], which dissolves very easily in ethanol and works (bonds) down to the lowest temperatures. If the investigated material does not decompose or change its properties in ethanol, the investigated substance can be stabilized in the GE varnish and be affixed firmly to common sample holders, like rectangular pieces of semiconducting compound, and then mounted into plastic straws [32] or deposited on Si [25,33,34]. In all other cases, different, usually custom devised approaches had to be developed [35]. As far as the present authors can judge, none of them can be regarded as magnetically transparent or being able to provide truly quantitative results.

In this context, small capsules appear to be the solution of choice in magnetometry of small quantities of shapeless materials, which evade other known methods of sample mounting for the measurements [36,37]. There are many varieties of capsules available, but for the magnetometry the capsules should be first of all small, to minimize their own magnetic contribution to the research, and colorless (transparent), to minimize the amount of dyes that usually add a paramagnetic contribution to their generally diamagnetic response. These and other contaminants make their magnetic properties difficult to standardize. The second serious issue related to the capsule-based magnetometry is the lack of simple and magnetically clean means to firmly fix the specimen-containing capsule to the magnetometer sample holder. By far the most popular Quantum Design (QD) [38] lines of magnetometers—MPMS-XL and the MPMS-3 VSM SQUID—have provided means to use transparent drinking straws for sample housing. Indeed, most of the SQUID magnetometry research has been performed with an aid of these straws [20,39,40]; other, more clean but elaborate solutions are exercised less often [19,22].

It turns out the inner diameter of the straws provided by QD exceeds the outer diameter of capsules that can be obtained from the same source. Neither size 4 gelatin capsules (GC) nor #5 polycarbonate capsules (PCC) can be stabilized firmly within these straws, so various means of their stabilization have been exercised. Most of these ad hoc solutions, like indentations of the straw below and above the capsule [37], insertion of pieces of cotton wool, various plastic sticks, or short pieces of somehow deformed straws might have proven sufficient for the fixing problem, but each of them introduces their own magnetic contamination to the measurement process. Therefore, these solutions work very well only when the researched signals are strong, exceeding at least by an order magnitude the total magnetic signal of the capsule and its harness within the straw. However, frequently, the researched material is characterized by a weak magnetic response, say of a range typical for common diamagnetic substances, and/or for various technological reasons only miniscule amounts are available for the studies. Therefore, a new approach to precision integral magnetometry is needed to give a better experimental account in these situations where the investigated signals are far smaller than that of the capsule and its surroundings.

In this report we provide a complete solution to tackle the challenges related to precision integral magnetometry of miniscule amounts of shapeless substances. The method we put forward here combines both the capsules and modified plastic straws in a magnetically clean manner that allows the determination of the absolute magnitude of the magnetic moment of single-milligram-small diamagnetic samples with 1% accuracy. The method is validated for a crystalline body and exemplified for a powder sample. It has clear potential to considerably impact the research of these liquid solutions and wet biological specimens that will withstand the dry environment of the SQUID sample chamber without marked modifications to their properties. The latter line of research has not yet been exercised by the authors.

Our report is organized as follows. We first introduce the method of permanent plastic straw modification that produces straws with customized (reduced) internal diameter. This makes them suitable to accommodate capsules and provides sufficiently strong grip permitting 4 cm long scans in 1 Hz oscillating mode of the fast averaging mode, the so-called reciprocal space option (RSO) of the MPMS systems. We then introduce the method of *in situ* compensation of the signal of the specimen-housing capsule by substantially restoring the symmetry of the specimen holder that is otherwise broken by the presence of the capsule. We provide the complete record of the assembly process of the capsule compensational sample holder (C-CSH) and show that a 30-fold reduction of the apparent mass of the capsule, that is, the mass of the capsule that is sensed by the SQUID, is readily possible. We finally validate our method by measuring the complete temperature dependence of the magnetic susceptibility of a 2.1 mg small crystallite of diamagnetic SnTe in our C-CSH. Other measurements of a 2.9 mg sample of common turmeric powder exemplifies the applicability of our approach to investigations of equally small amounts of powders, which we quantify numerically. All of these accomplishments are done using only data reduction that is based on the standard output files of the MPMS system (*.dat files). It is shown that none of the excessive post-measurement data treatment is necessary. The method outlined here has been proven to be suitable for measurement from 2 to approximately 330 K and in the whole range of the magnetic fields available within the magnetometer and to be compatible with the RSO mode of measurements.

**2. Materials and Methods**

The main purpose of this report is to provide the reader with the complete method developed to *in situ* compensate for a large magnetic contribution brought about by common capsules, frequently employed to house soft matter or just shapeless specimens, which are difficult to be fixed to a reasonable sample holder for the intended magnetic studies. Therefore, in this section we stipulate only that the main building blocks of the self-compensating assembly are polycarbonate or gelatin capsules arranged in trains with (nearly) identical neighbors, which are all fixed into thermoplastically modified plastic drinking straws. All these elements are available directly from Quantum Design (San Diego, CA 92121, USA), and we advise to use this source for the sake of standardization. Gelatin capsules (GC) of size 4, or #4, are provided as item QDS-AGC1. The polycarbonate (PCC) ones of size 5 (smaller), or #5, can be ordered as item QDS-AGC3. Both are packed in bunches of 100. However, the inner diameter of the straws provided by QD (about 5.3 mm) exceeds the outer diameter of standard capsules, which are about 4.6 and 5 mm for PCC #5 and GC #4, respectively.

*2.1. The Capsules*

Apart from the mismatch of the external diameters of capsules with the inner diameter of straw, which determines the amount of effort required to fix them to a sample holder, there are other important factors that need to be discussed here. These are: (i) their magnetic susceptibility, which determines the magnitude of the magnetic response or the signal brought by the capsule to the total magnetic moment *m* registered during the measurement, (ii) their mass, which quantifies the weight of the sample that can be housed in the given capsule to produce an easy distinguishable magnetic response, and (iii) alignment (proper centering) issues of small specimens. The last factor is somehow the derivative of the former two. In these respects, there are virtually no problems or constraints for investigation of ferromagnetic materials of which even nanograms can dwarf the response of the capsule, particularly at a weak magnetic field. The method of compensation we put forward here will be certainly suitable for such samples, but in the report, we put our emphasis on weakly magnetic substances that are characterized by the weakest magnetic responses—the diamagnetic bodies. Our effort has been directed to minimize as much as possible the response of the capsule, so as to bring forward the weak or even very weak response of the specimen and to quantify its magnitude with a very high accuracy.

Both types of capsules have different sizes and masses. The #5 PCCs range between 22 and 35 mg and the #4 GCs between 30 and 44 mg, but it turns out that both types are characterized by a similar mass susceptibility $\chi$ of about $-6.3(2) \times 10^{-7}$ emu/g/Oe. Therefore, for the sake of simplicity and to relate the results to the similarly diamagnetic specimens of a very small mass, wherever possible, the magnetic results are converted to the mass of an idealized, yet quite typical, mass of polycarbonate capsules, i.e., to 30 mg. In that sense this conversion will make the conclusion of our findings independent of the magnitude of the magnetic field *H*, at which the measurements are taken, as both signals of the capsule(s) and the specimen scale similarly with *H*. Additionally, for the sake of argument, we define a standard reference diamagnetic material (SRDM) and assign to it magnetic susceptibility $\chi_{SRDM} = -5 \times 10^{-7}$ emu/g/Oe, that is, a kind of broadly understood average between room tempera-ture magnitudes of $\chi$ of capsules and other commonly met materials as: Si ($\chi_{Si} \cong -1.2 \times 10^{-7}$ emu/g/Oe), GaAs ($\chi_{GaAs} \cong -2.25 \times 10^{-7}$ emu/g/Oe) [41], or sapphire ($\chi_{sapphire} \cong -3.5 \times 10^{-7}$ emu/g/Oe) [17].

2.1.1. Hygroscopicity of Gelatin Capsules—Time Domain Studies

We decided to standardize the magnetic results to the mass of typical polycarbonate capsules because they can be regarded as stable in ambient conditions, in stark contrast to the gelatin ones, which are highly hygroscopic. As the result, they are changing their mass in response to the surrounding humidity. The effect is particularly strong after insertion a typical GC to the SQUID chamber. In Figure 1, we compare time dependencies of $m$ of both types of capsules. The effects seen for the GC are paramount, its time dependence of $m$ is very substantial. There are some basic features which need to be highlighted:

1. Within 20 h spent in the dry residual helium atmosphere in the SQUID sample chamber, the capsule lost nearly 10% of its magnetic response. This ties precisely with the equivalent loss of about 3 mg of its weight—recorded accordingly at the key points of this investigation (the numbers are given in the caption). The equivalent mass of the capsule is indicated on the right Y-scale. This magnitude of change of mass is 50% greater than the mass of the sample that is used for the validation of our C-CSH presented further in the report;
2. Two-thirds of this change takes place within the first 3 h. In other words, the change of the signal during the typical time needed for the measurement of a standard $m(H)$ hysteresis curve corresponds to the full mass (that is to the total signal) of about 2 mg of SRDM;
3. Upon removing the sample from the SQUID for about 2 h, the capsule regains approximately 0.8 mg of its weight, but when reinserted to the chamber, the process restarts again and continues with a similar pace. We assign these environmental changes to desorption and absorption of water vapor;
4. The same test executed for a polycarbonate capsule indicated only a very weak gain in $m$ (an equivalent of less than 0.2 mg of its mass) seen within first 20 min of the test. This process saturates swiftly and no $m$ dependence on time is subsequently detected, as expected for a polycarbonate material.

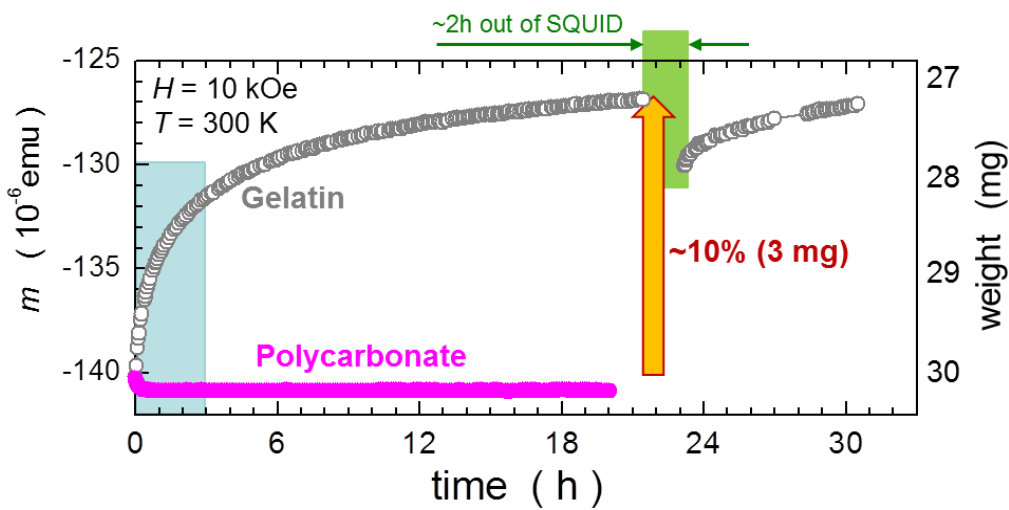

**Figure 1.** Comparison of time dependencies of the magnetic moment $m$ of typical gelatin (grey symbols) and polycarbonate (magenta) capsules. All magnetic data are normalized to the standard 30 mg capsule and the equivalent scale of mass is given on the right Y-scale of the graph, established from weights of the capsules taken just before inserting to and just after removing from the SQUID sample chamber. The green band in the background indicates the time the gelatin capsule spent in open air during a break in the measurements. The light-blue band at the beginning of the graph marks the typical time needed for the execution of a standard hysteresis curve. The capsules are harness-free mounted in a diameter-reduced straw, as explained in Section 2.2.

2.1.2. Temperature Dependence

The temperature dependent studies of single capsules are summarized in Figure 2. Similar to the Figure 1, we scale $m$ of GC to that of PCC and both relate to the standard mass of 30 mg. Importantly, for this measurement the GC has been kept in the pumped SQUID chamber for nearly 5 h to somehow saturate the dehydration of the gelatin. There are two major features of these results:

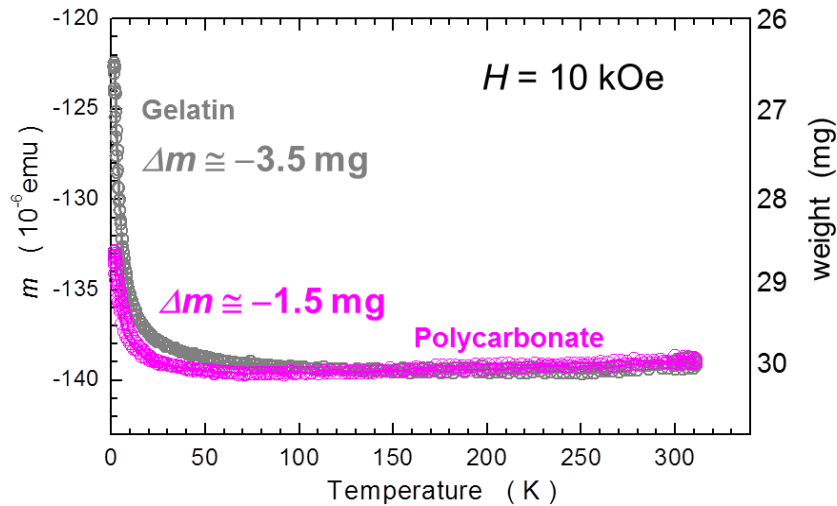

**Figure 2.** Temperature dependence of the magnetic moment $m$ of single gelatin (grey) and polycarbonate (magenta) capsules. Magnetic data are normalized between themselves and a related standard mass of 30 mg. The magnitude of the paramagnetic shift of $m$ at low temperatures is expressed as an equivalent "loss of mass" of the capsules, $\Delta m$. The capsules are harness-free mounted in a diameter-reduced straw, as explained in Section 2.2.

1. The temperature induced changes are relatively small down to about 50 K and relate to about 0.2 mg of SRDM. This is an encouraging result; however, the gelatin capsule had to be seasoned in the SQUID for about 5 h before the $T$-scan commenced;
2. Below about 50 K, strong paramagnetic contributions occur, which corrupt the otherwise ideal diamagnetism of the capsules. It is instructive to relate the reduction of the magnetic response of the capsules to an apparent change of mass of a sample corresponding to SRDM. We find that this change is as strong as above 3 mg for GCs, and is approximately half the magnitude in PCCs. This finding puts a strong constraint on precision magnetometry at low temperatures of miniscule samples requiring encapsulation. Importantly, this effect is practically impossible to compensate for due to unavoidable misalignment errors, discussed in the next section. The only positive aspect of this feature is that it is fully reversible with $T$.

In the main section of this report we show that all these detrimental properties to precision magnetometry of miniscule specimens can be substantially reduced—more than ten times—that is, it can be brought to the level corresponding to 0.2 mg of SRDM.

2.1.3. Capsules as Containers for SQUID Magnetometry (Centering Issues)

The factor that practically precludes precision integral magnetometry using capsules as housing carriers of miniscule samples of a very small $m$ is the challenge of the adequate centering of the whole sample holder (alignment) and related inability of the extraction of the credible magnitudes of $m$ of the scrutinized specimen. We remark first that there is no centering problem when the investigated material fills the whole volume of the capsule, as exemplified by the first four capsules in Figure 3. The magnetic moment of the sample is distributed evenly along the length of the sample and the MultiVu centering routine will direct the user to the proper alignment, marked as the black dashed line. There are also no big issues with establishing the accurate magnitude of *m* of such massive amounts of material. The reference measurement of the empty capsule will provide the correct magnitudes of the background signal. Alternatively, the studies can be carried out within the automatic background subtraction (ABS) method frame.

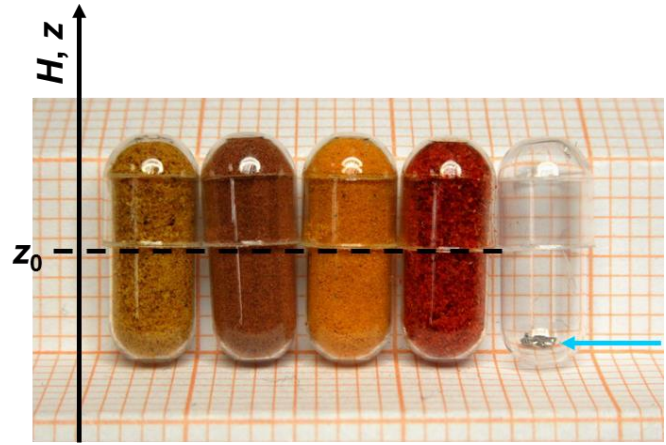

**Figure 3.** Some examples of the use of small capsules (#5 in this example) for measurements of substances that otherwise would be difficult to reliably and cleanly mount in the SQUID sample chamber. From the left, the first four are fully loaded, so no alignment problems are expected. The magnetic "center of gravity", marked by the dashed black line, coincides with that of the capsules. These capsules house common curry, cinnamon, turmeric, and red pepper powders. The fifth capsule houses a small, 2.1 mg, crystallite of SnTe, used later to validate the *in situ* compensation method put forward in this report. As indicated, the crystallite sitting at the bottom of the capsule, is shifted by about 5 mm from the center of the capsule, causing serious experimental errors due to alignment problems, discussed in this section.

The problems begin when a small specimen with a weak magnetic response occupies only the bottom end of the capsule, as exemplified by the fifth capsule in Figure 3. Such an object will cause only a small deviation to the dominant response of the capsule. Certainly, one can use a custom-made magnetic marker attached to the bottom of the capsule (Appendix A) to center the sample holder at this location. There is, however, a limited accuracy, or reproducibility in placing the marker, so the alignment errors are unavoidable. We evaluate the misalignment errors by multiple fitting of the SQUID response function (SRF, Appendix B) to the typical scan of a standard capsule, each time setting a different center position $z_0$ for SRF. The resulting dependence of *m* on $z_0$ is depicted in Figure 4 by pink bullets. For convenience, we express the magnitude of *m* by the corresponding weight. As can be seen, *m* drops sizably when we move away from the center of the capsule, $z_0$ = 2.0 cm. More informative is the derivative of this dependency, as it directly quantifies the magnitude of the misalignment errors. We read that it is as high as 10% of the mass of the capsule or simply about 3 mg of the SRDM per one millimeter of the misalignment, when one wants to extract the moment located at the vicinity of the bottom of the capsule. Clearly, only strong magnetic materials can be accurately investigated in such a manner.

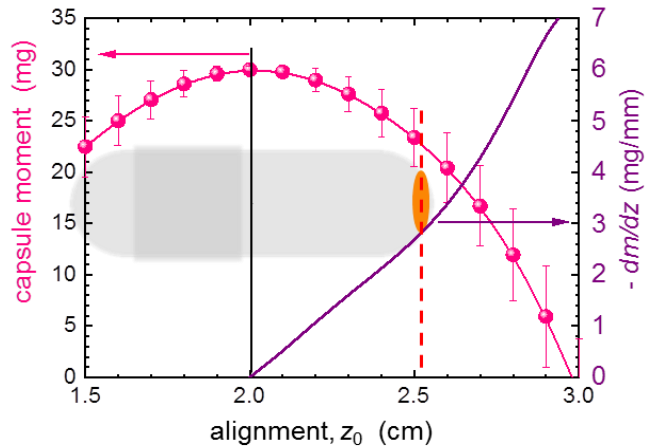

**Figure 4.** The position dependence of the magnetic moment calculated from a scan of a typical #5 capsule (30 mg of mass) expressed as the corresponding mass (pink bullets). The magnitude of this change defines the absolute value of the misalignment error with respect to the center position (a black vertical line at $z_0 = 2.0$ cm). The purple solid line yields the derivative of the former, that is the magnitude of the uncertainty due to the misalignment. The sketch of a #5 capsule at the background brings a sense of the real scale. The orange oval mimics a small object of investigation residing at the bottom of the capsule at the position marked by a dashed red line.

*2.2. Reduction of the Diameter of the Straws*

The clear drinking straws can be regarded as the main workhorses of the commercial integral magnetometers due to their "stealthy" properties. Interestingly, these properties are not related to their negligible weak magnetic properties. On the contrary, the magnetic susceptibility of the material of the straw is quite high for a diamagnetic substance: $\chi_{straw} \cong 8.8 \times 10^{-7}$ emu/g/Oe. At least, such a number has been established at room temperature for a 5 mm long piece of a straw coming from the same batch of straws used in this study. This is some 60–70% more than that of the capsules and of SRDM.

The straws earned their reasonably good reputation [20,32] as they are sufficiently homogeneous and long, so that during the scanning movement no appreciable changes of the magnetic flux are sensed by the detection circuit of the SQUID. The same straws constitute therefore the basic scaffolding within which the *in situ* compensation assembly is nested. However, the typical internal diameter of the straw is about 5.2 mm, so it exceeds the external diameter of the either type of capsules considered for the use in the magnetometric effort. The capsules move freely along the whole length of the unmodified straw. Since it is hard to design a magnetically stealthy harness for the stabilization of their position in the straw, we present a method for adequate reduction of the diameter of the straw that allows for precise, reproducible, and above all, harness-free firm fixing of capsules. Most conveniently, the design we put forward withstands even very long measurement runs performed in the oscillatory RSO mode without any noticeable change of its alignment with respect to the sensing coils, so no changes to the everyday experimental routines are necessary.

The key factor allowing the required modification of the diameter of the straws is their thermoplasticity. The straws become plastic at approximately 80–100 °C, and only if they are cooled in the adequate mold do they preserve the shape imposed by that mold, at least until reheated back to be plastic again. As the capsules have their external diameter smaller than the straws, we reduce the diameter of the straws by the following procedure:

- We select four straight straws that we clean thoroughly in ethanol. One of them is left aside;
- Next, we incise the wall of the remaining three straws along almost their whole length. That is, the cutting starts approximately 6–8 mm from one end and in one steady move is continued until the other end of the straw is reached. A sharp and clean scalpel blade should be used. The incision should be as smooth as pos-sible, torn edges should be avoided—they are likely sources of additional magnetic signals, which makes such a straw useless as the sample holder;

- We now insert the first of these three cut straws into the uncut one. Obviously, despite being cut along their length, the cut straws preserve their original diameter. Therefore, to nest a cut straw into the uncut one, the inserted straw has to curl a bit to reduce its external diameter to the internal diameter of the uncut one. We press the straw as far as it goes. It stops where the incision begins;
- We continue the process with the second straw. This one has to curl more to nest into the reduced clearance of the already joined first two straws;
- The third inserted straw has to curl even more, so a noticeable resistance can be felt. One can skip this point if serious problems occur. Actually, it is the second straw that we aim for. The reduction of its diameter is sufficient to firmly hold the #5 PCC capsule. All four (three) combined straws should be now heated to approximately 80 °C. Any laboratory heater will do. At this temperature, the material softens and the straws are re-formed to their new diameter set by the outer straw. In this sense the external uncut straw serves as a mold for the whole stack. The whole assembly ready for baking is depicted in Figure 5;
- We typically allow the straws to stay there for about half an hour, after which we let them to naturally cool down to room temperature. After the cool down, the straws can be separated. Three (two) straws with new diameters has been formed. We mark them as S0 (the uncut one) and consecutively, S1, S2, and S3. We confirm upon multiple tests that neither the smooth incision nor the baking and shape modifications introduce any noticeable modification to the magnetic homogeneity of the straws.

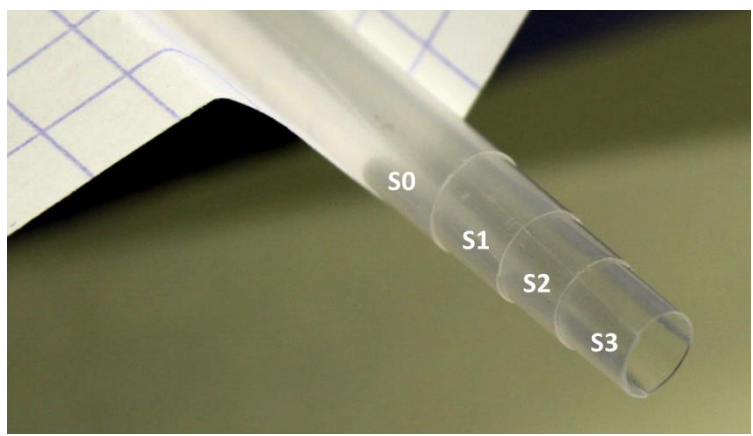

**Figure 5.** A complete set of three cut straws (S1, S2 and S3) inserted into the original straw S0. To nest in S0, the next straws have to progressively reduce their diameter by adequate curling.

The thickness of the wall of the straw is about 0.15 mm. At the first glance this may indicate that straw S1 has its internal diameter reduced by 0.3 mm, S2 by 0.6 mm, and so on. However, the clearance of straw S1 is smaller due to the overlap of the walls along the incision. Therefore, its clearance is reduced by three thicknesses of the wall, that is by 0.45 mm, to approximately 4.8 mm. Straw S2 has to curl even more in S0 and S1, so its clearance is further reduced to approximately 4.3 mm. We note that this is already less than the external diameter of the PCC #5, and straw S2 is found to be the best suitable to house these capsules.

The main usefulness of the modification described above stems from the existence of the incision along the nearly whole length of the straws. This permits the insertion of object(s) larger than the current clearance of the straw. The walls of the cut straw step apart in the vicinity of the object(s) to locally increase the clearance. Effectively a bulge is formed, as exemplified in Figure 6. Within the bulge, the walls of the straw press against the object, say a capsule (but it can be a rectangular sample of other crystalline material), and the friction between the walls and the objects holds the object firmly in place. The grip is so sufficient that the authors have never experienced any shift of the position of a #5 capsule in an S2 straw, even being loaded up to 180 mg. The meas-urements of the empty capsules presented in Figures 1 and 2 were performed having the investigated capsules clamped by the expanding walls of an S2 straw. The same straw accommodates #4 capsules, except the bulge gets larger.

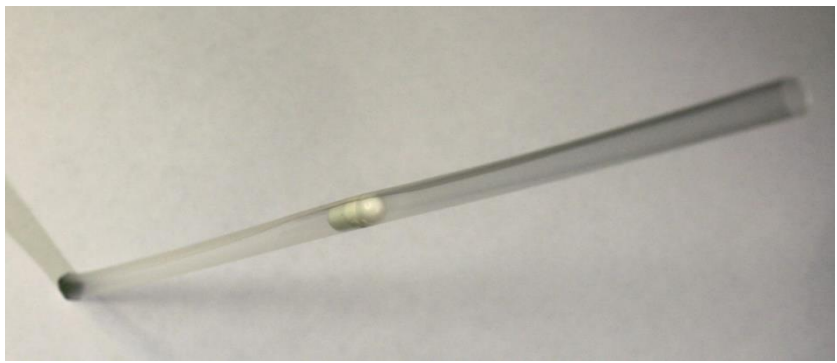

**Figure 6.** An illustration of a bulge formed on an S2 straw with its diameter reduced below the external diameter of a capsule #5 (for the sake of illustration a white coloured one is used). The S2 straw bulges easily due to the incision running along its length. The same incision allows to effortlessly adjust the position of the capsule, whereas the clamping force of the walls of the straw assures its alignment even during the reciprocal space option oscillatory movement in the sample chamber.

It is a matter of trial and error to find the best configuration. The authors have tested that straw S2 flawlessly holds a #5 capsule filled up to 150 mg during 20 h long 1.5 Hz and 4 cm long RSO measurement. For larger diameters, straw S1 should be used. It can house any capsule from #4 to #1. The latter has the external diameter of approximately 6.9 mm, however, in this case we suggest using a second S1 straw wrapped around the first one to increase the clamping grip.

We summarize this part by emphasizing that the single incision along the nearly entire length of the straw allows:

- A permanent reduction of its diameter (clearance) during an adequate heat treatment;
- Practically unrestricted movement along the straw of objects with effective diameters exceeding the modified diameter of the straw;
- A position-stabilizing grip sufficient to withstand acceleration and deceleration forces during the RSO measurements.

*2.3. Magnetic Measurements*

All measurements in the report have been taken using two commercial MPMS-XL magnetometers operating to a maximum field of $H_{max}$ = 50 and 70 kOe, located at the Institute of Physics, Polish Academy of Sciences in Warsaw. Therefore, most of the technical remarks are strictly applicable to these units. However, the considerations presented below are so general that the required adaptive modifications could be employed to sizably improve the credibility of the research studies in other types of magnetometers, such as the SX700 of Cryogenic Ltd. [42] and the more popular SQUID-VSM MPMS3 of Quantum Design.

One of the main advantages of the *in situ* compensation method described here is that it is fully compatible with the RSO measurement mode, embedded in the MPMS-XL systems. The method has been derived to work with it, tested and validated using this improved oscillatory (mechanical) method of scanning. It allows for the collection of more scans in a given unit of time, which after the averaging procedure provide higher sensitivity and a reasonable sense of the experimental error. All the data presented here are collected by setting the scan length to 4 cm and RSO frequency to 1 Hz. Six measurements each consisting of four oscillations (24 scans all together to be averaged) are used to define the magnitude of each single experimental point. Further consideration on these setting can be found in Appendix C. Both the linear and iterative regression modes of the fitting routine (extraction method of the magnitude of the magnetic moment *m*) are used depending on the specific requirements. We use the sweep approach for the temperature *T* dependent studies.

## 3. The Capsule Compensation Sample Holder

The step-by-step assembly of the capsule compensated sample holder (C-CSH) is described here. Importantly, the description will provide a very good sense of the principle which makes the C-CSH so effective, robust, and easy to work with. We detail the process for the #5 polycarbonate capsules.

*3.1. General Considerations*

The construction process should be preceded by a careful selection of the capsules. Only capsules with very similar masses should be used. PCCs are provided by QD in two plastic bags containing separately the base parts (the containers) and the caps. This is an initial advantage as they exhibit a sizable mass distribution. The masses of the container parts range from 18 to 25 mg; the masses of caps range between 12 to 16 mg. As such, the distribution of masses of randomly assembled capsules may indeed be substantial. It therefore pays to spend the time weighing at least 30 containers and caps and arrange at least 15 capsules with the least possible spread of weights. We recommend to punch a tiny hole on the rounded part of the cap of each capsule. We use a needle from an insulin injection syringe. The hole will increase the effectiveness of the air evacuation from the interior of the capsules during purging of the sample chamber of the magnetometer. Without this hole, a tightly fitted cap can cause the capsule to slide-open when the drop of the pressure inside the capsule is much slower than in the sample chamber.

Having the weights organized, the user can start loading the capsules from any end to the already prepared straw S2, one after another, or the loading can be done as in the example detailed below. The already assembled train of 15 PCC capsules is depicted in Figure 7. The figure also documents our selection effort to prepare the set of several capsules of nearly identical weights.

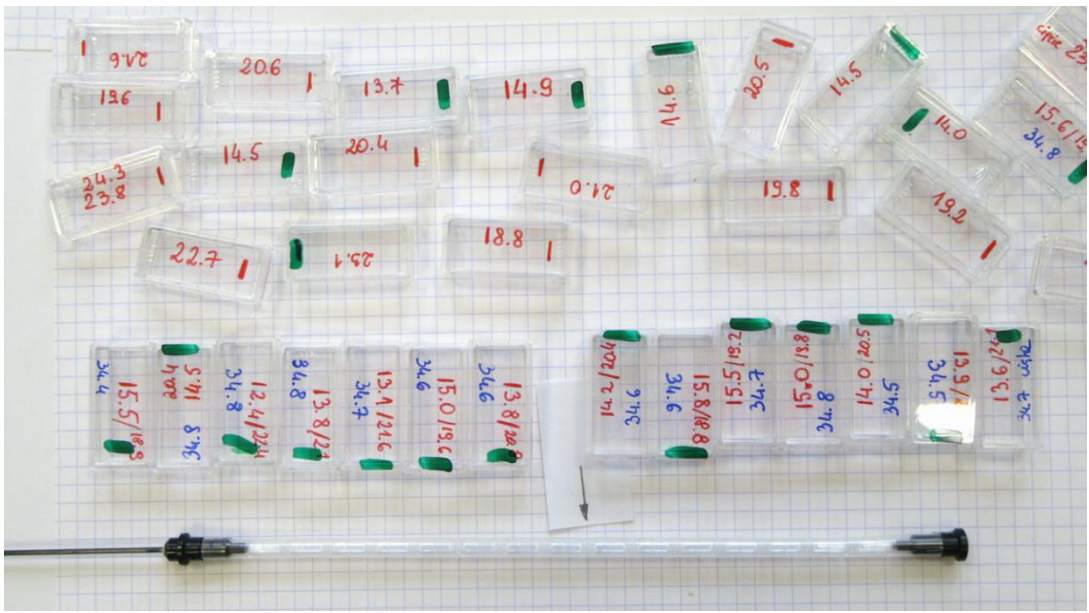

**Figure 7.** (**Bottom**) Fully assembled capsule compensated sample holder (C-CSH) made from 15 #5 polycarbonate capsules tightly combined in an S2 straw to make the self-compensated train. The arrow marks the position in which a small object will reside during the measurements. (**Top**) The markings on the plastic boxes located just above the C-CSH state the weight of each capsule used to assembly the train. Other boxes at the upper half of the figure are the leftovers from the capsule selection process.

*3.2. The Assembly of C-CSH and the Principle of Its Effectivness*

We start the assembly from placing the central capsule, the one in which the specimen will be located, at the center of the straw selected to carry the whole ensemble, as presented in Figure 8a. For the sake of presentation, we illustrate the assembly process using white-colored #5 gelatin capsules. The results of the SQUID measurements presented on the right side of Figure 8 are taken for the transparent PCCs discussed in this study, as presented in Figure 7. All of these measurements are performed at room temperature and at $H$ = 10 kOe. The corresponding scan (the "Long Avg. Scaled Response", see Appendix B) registered by the SQUID is marked in Figure 8b as grey symbols. This scan is our reference point. It exhibits the typical shape, distorted noticeably from the ideal point object response because of the extended nature of the capsule with its length of about 11 mm. We calculate the corresponding magnetic moment using the standard point object SQUID response function (SRF). The technical aspects of this process are detailed in Appendix B. We consider two cases. First, the center of the scan window is fixed at 2.0 cm, which is at the center of the capsule. This result is marked by black line in Figure 8b. Second, the center of SRF is fixed at 2.5 cm—corresponding to the bottom of the capsule. The result of this fitting is marked by red line in Figure 8b. The former case represents the capsule completely filled with the researched material; the latter one mimics the situation when the capsule contains only a miniscule amount of the investigated material, which resides at the bottom. The obtained magnitudes of *m* are depicted in Figure 8d by gray-filled circles, with the edge color (black/red) corresponding to the position at which we fix the center of the SRF.

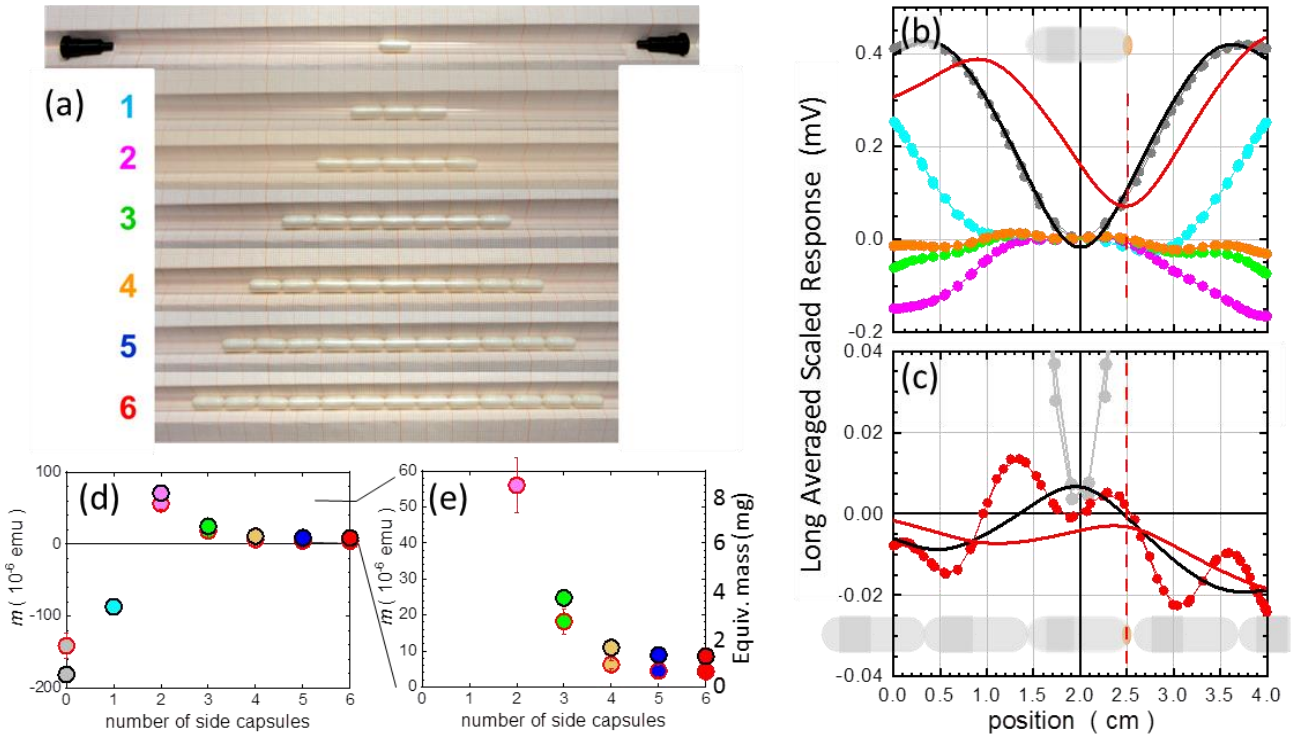

**Figure 8.** (**a**) The illustration of the step-by-step assembly (subpanels 1–6) of an *in situ* capsule compensational sample holder from capsules of size 5 and an S2 straw, that is, a straw with its diameter reduced below the external diameter of the capsules. White-colored #5 gelatin capsules are used for increased visibility only. Results presented in the rest of the panels are obtained for transparent polycrystalline capsules, as presented in Figure 7. (**b**,**c**) SQUID scans recorded for a single central capsule (grey symbols) and 1st, 2nd, 3rd, 4th, and 6th stages of the assembly (symbols of matching colors to the numerical markers in panel (**a**)). The black vertical line represents the position of the midpoint of the capsule in the scanning window of SQUID magnetometer and the red dashed one the position of the bottom of the capsule. The black and red wavy curves represent the SQUID response function (SRF) centered at these two locations. In panel (**b**) the fitting is done to the dataset

corresponding to a single capsule, in panel (**c**) to the 6th stage of the assembly (red symbols). The fitting approach is described in Appendix B. (**d**) Results of the fits to the scans collected in every stage of the assembly (magnetic moment *m*). The fill color of each circle corresponds to the assembly stage, whereas the edge color (black or red) correspond to color marking of the alignment of the SRF: at the center of the central capsule or at its bottom, respectively. (**e**) The blown-up part of panel (**d**) for increased resolution near zero for the last stages. The added right Y-scale expresses the degree of the compensation in an equivalent mass of the central capsule of a mass of 30 mg.

The assembly continues by addition of one capsule on both sides of the central capsule, as shown in Figure 8a, case "1". The magnetometer registers this change. The form of the scan corresponding to this configuration (cyan symbol in Figure 8b) assumes a trapezoidal form and the overall amplitude is visible reduced. This is immediately reflected in the calculated magnetic moments corresponding to this configuration (cyan filled circles in Figure 8d). The main message here is that the detection unit of the magnetometer now reports a twice smaller magnetic moment despite having three times more material in the chamber. The improvement continues after placing next neighbors on each side, Figure 8a, case "2", and magenta filled circles in Figure 8d. However, an interesting thing happens now: the scan inverts (magenta symbols in Figure 8b), so the train of five diamagnetic capsules is "seen" as a blurred "paramagnetic" object.

A far more substantial change is observed when there are four compensating capsules on each side of the central one, Figure 8a, case "4", that is, when the total length of the train of capsules (about 10 cm) guarantees that neither end of the train gets very close to the pick-up coils during scanning. This makes our assembly semi-infinitely long, fulfilling the first prerequisite condition of the efficient *in situ* compensation. Indeed, the voltage registered during scanning is strongly reduced (orange symbols in Figure 8b), particularly in comparison to the initial case of the single capsule. The corresponding magnetic moments are marked as orange filled circles in Figure 8d. They are now very close to zero, but not strictly there. First, the train is not really very long (yet); second, it is not uniform due to its "granular" form. The capsules are not uniform themselves, they are thicker at the cap-containing half and are rounded at both ends. The resulting nonuniform mass distribution along the length of the train of capsules will always have its mark on the final form of the scan. It will be shown to be appreciably small.

We continue the assembly by adding two more capsules to each side. Actually, only five compensating gelatin #4 capsules can be added on each side of the central one. Capsules #4 are so long that 11 of them nearly fill the whole length of the straw. In the case of PCC capsules #5, one can go up to 7 capsules on each side to achieve additional positional stability with respect to the straw, and so to the SQUID pick-up coils—fulfilling yet another prerequisite condition for achieving very high sensitivity and reproducibility. The fully assembled train of 15 PCC capsules is depicted at the bottom of Figure 7, together with the documentation of the selection effort to prepare the set of several capsules of nearly identical weights. The shape of the scan obtained for the fully assembled train is marked by red bullets in Figure 8c. It certainly is very "bumpy" on its own scale, but its magnitude is truly small compared to the response of a single capsule. The SRF fitted with its centering forced to our two points of interest (black and red solid curves) yields magnetic moments indicated by black and red points in Figure 8d, and in its blown-up version, Figure 8e. By comparing these results with the starting ones for the single capsule, we obtain that magnetic signature of the *in situ* compensated capsule dropped about 20 times (nearly 95% of compensation) when one seeks the response at the location corresponding to the center of the capsule and about 30 times (nearly 97% of compensation) when the response at the location of the bottom of the capsule counts most. To visualize the effectiveness of the *in situ* compensation, the magnetic moment is again expressed in units of mass, the right scale of the graph in Figure 8e. These numbers read approximately 1.5 and 1 mg, respectively. This fact alone indicates that magnetic moments of a SRDM can be established in C-CSH with an *absolute* accuracy of 10 to 20% when the mass of the investigated specimen ranges between 10 and 5 mg, respectively. It is shown later in Section 4: "Discussion: validation and an example of the true potential" that one can get a much better accuracy when this residual contribution from the compensated capsules (the empty C-CSH) is considered.

Another advantage of the approach described here is that the standard mounting method of the C-CSH assembly with the long and short sample rods, required for the RSO method of measurements, can be applied. At least one black centering plug (straw adapter) has to be used to connect the straw with the rest of the sample rod. This is the reason why the incision on the straws starts a few millimeters from their one end. Here enters the connecting plug. And our experience is that for mechanical stability of the whole C-CSH assembly, one plug is acceptable. Straw S2 provides a sufficient grip to hold the train of a dozen or so #5 capsules firmly in place. However, the second plug is strongly advisable to assure a proper centering of the straw with capsules with respect to the axis of the SQUID pick-up coils. At the second (the lower) end the straw is cut and here the grip is insufficient to hold the plug. To secure the centering plug in place, we apply a small patch of a commercial surgical tape with acrylic adhesive. We elaborate a bit more on this very handy fixing agent in Appendix A.

Finally, we want to underline the usefulness of using diameter-modified straws for C-CSH-type assemblies. The key point is that the larger internal diameter of unmodified straws will allow the capsules to tilt or even rattle during the RSO measurements, an effect most likely more substantial at very low temperatures. This may add up to misalignment errors and thermal hysteresis of the results. The authors find the existing friction between the diameter-reduced straw and the capsules as advantageous during assembly and disassembly of the C-CSH. Only the bottom half of the train has to be removed from the straw to get to the central capsule to exchange the specimen. The rest of the assembly retains its location. This greatly increases the reproducibility of the assembly and so minimizes the probability of a sizable misalignment. The wall-wall friction between capsules and a straw is even more crucial in the case of gelatin capsules. As noted in Section 2.1, GCs are changing their mass when they experience the dry environment of the SQUID sample chamber (Figure 1). Similarly, their linear dimensions are changed. In particular, they shrink visibly, and so after some 10 h long measurement run an 18 cm long train of GCs gets 2–3 mm shorter. Without the radial grip provided by the wall of the diameter-reduced straw, such a contraction would result in a time-dependent 1–1.5 mm alignment shift of the specimen in the central capsule.

*3.3. Examples of the Improvement*

By extending the "length of the capsule" in such a harness-free manner, the capsule(s) become "stealthy". The improvement brought about by the *in situ* compensation is not only related to the fact that the effective mass of the capsule that houses the specimen is reduced so sizably; the gain is multiple.

At first, we note that the concept of *in situ* compensation makes the outcome of the experiment much less dependent on the misalignment error, that important and detrimental factor when the background measurement has to be performed and subtracted. We illustrate the improvement by performing similar $z_0$-dependent fitting as summarized previously in Figure 4. However, this time the SRF is fitted to the to the exemplary SQUID scan of the fully assembled empty C-CSH, red symbols in Figure 8c. The results are depicted in Figure 9, clearly indicating that misalignment errors dropped 10 to 15 times with respect to the case of a single uncompensated capsule, and that a substantial misalignment of the about 1 mm should not introduce a larger experimental uncertainty than 0.1 mg, or 1% during investigation of 10 mg of SRDM.

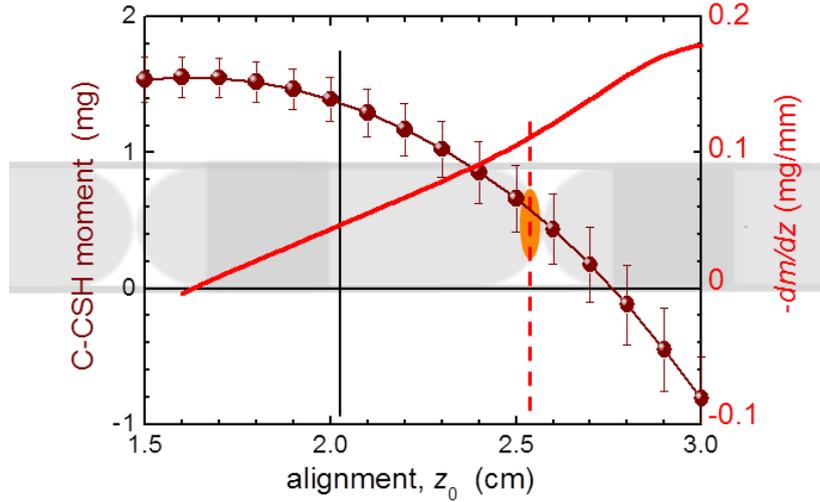

**Figure 9.** The position dependence of the magnetic moment (brown bullets) calculated from a scan of a fully *in situ* compensated train of #5 polycarbonate capsules (sketched to scale at the background) forming the capsule compensation sample holder (C-CSH) assembly. The results are related to the standard mass of a single capsule (30 mg, as in Figure 4). The magnitude of this change defines the absolute value of the misalignment error with respect to the center position (a black vertical line at $z_0$ = 2.0 cm). The red solid line yields the derivative of the former, that is, the magnitude of the uncertainty due to the misalignment. The orange oval mimics a small object of investigations residing at the bottom of the central capsule at the position marked by a dash red line.

Anyone who is interested in very precise magnetometry of rather large quantities of shapeless materials, such as presented previously in Figure 3, when capsule is filled completely, can stop here. Indeed, the #5 capsule with its typical volume of about 0.130 mL can accommodate approximately 100–150 mg of lightweight material (or powder), an amount that exceeds the effective mass of the *in situ* compensated capsule by approximately 100 times. This guarantees at least 1% absolute accuracy without any background subtraction effort. However, it has to be immediately noted here that any number evaluated in such conditions has to be corrected for the extended shape of the specimen [19,43]. The correction factor for an 11 mm long object is about 0.85 [44], although the exact value is yet to be established. Therefore, the best magnitude of *m*, and so of the magnetization, can be obtained in this situation after dividing the output of the magnetometer (these number are computed in the point dipole approximation) by 0.85.

Certainly, the sizably increased accuracy of the *in situ* compensation is not the only benefit of the method. The self-compensation of the capsules aligned in a train also indicates that other disturbing features of the capsules presented in Section 2.1 are eradicated in a similar proportion. This is first indicated in the Figure 10, where we present the full *T*-dependence of *m* measured for two C-CSHs assembled from PCC #5 and GC #4. These results have to be related to the measurements of single capsules, presented in Figure 2. The most prominent improvement seen after assembling the capsules in trains is a perfect elimination of the strong low-temperature paramagnetic component. Interestingly, in both cases a small *T*-dependence is seen between 100 and 300 K, corresponding to about 0.2 mg of the SRDM. This value sets the lower limit on the precision of the method when one wants to work without subtraction of the background (empty C-CSH).

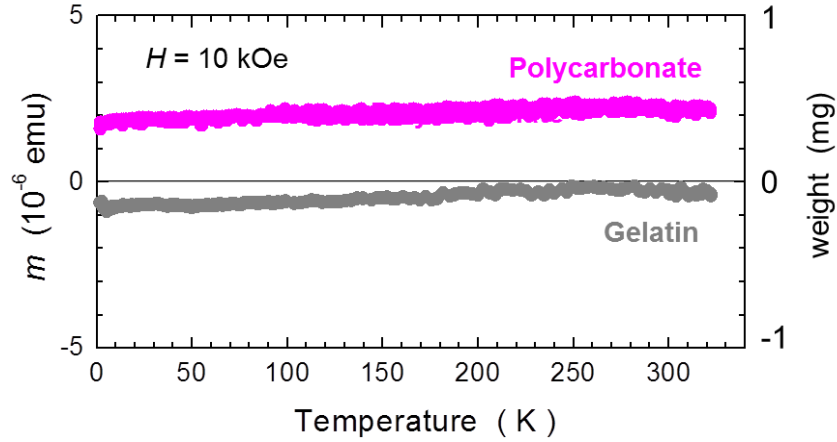

**Figure 10.** Temperature dependence of the magnetic moment *m* sensed at the location of the bottom of the central capsule for two exemplary capsule compensational sample holders assembled from gelatin (grey) and polycarbonate (magenta) capsules. Magnetic data are normalized between themselves and related to a standard mass of a 30 mg capsule. Note the complete lack of the strong low temperature upturn of *m*, the characteristic feature for single capsules (Figure 2).

All the test measurements are performed in $H$ = 10 kOe, indicating that the improvement has occurred in all three domains: time, temperature, and $H$. Nevertheless, it is worth presenting a direct measurement in $H$. In Figure 11, the $H$ dependence of the signal of single capsule is compared with that of a train of 14 PCC capsules, the C-CSH. In this particular case, the improvement is 45-fold, that is, the signal of the single PCC drops by so many times. Importantly, all the $m(H)$ isotherms for C-CSH show a very weak $T$-dependence, and particularly none at the lowest temperatures. Both $m(H)$ collected at 1.8 and 5 K go one on top of another. This confirms the results presented in Figure 10 and constitutes the base for the low-$T$ analysis of a paramagnetic contribution in turmeric powder, presented later in the text.

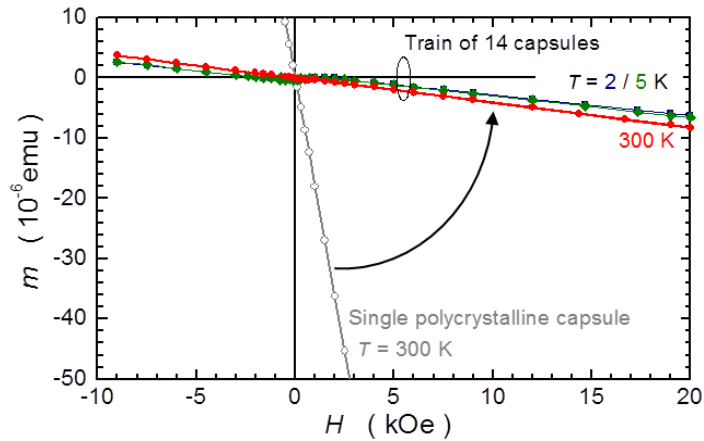

**Figure 11.** An example of the magnetic field $H$ dependence of magnetic moment *m* for a single polycrystalline capsule (grey circles) and of the *in situ* compensated case of a train of nearly identical capsules loaded and stabilised in a S2 straw having its diameter reduced below the external diameter of the capsules (navy, green, and red symbols). The arrow indicates the reduction of the magnetic signal of the capsule due to the compensation effect of the neighbouring capsules.

## 4. Discussion: Validation and an Example of the True Potential

In the previous section we introduced the whole concept of *in situ* magnetic compensation of the signal of common capsules that allow the investigation of various shapeless specimens in commercial SQUID magnetometers. In the following section, we will validate to method developed here to highlight its quite remarkable experimental potential.

In order to validate the method, we investigate a very small fragment of a crystalline material that is meant to mimic other specimens of a miniscule amount. An irregular 2.1 mg crystallite of SnTe is chosen. SnTe is a narrow gap semiconductor, the sole representative of the so called topological crystalline isolators [45]. The investigated material was grown using the self-selecting vapour growth method [46]. At room temperature its magnetic susceptibility is approximately $\chi_{SnTe} = -4.6 \times 10^{-7}$ emu/g/Oe (which ties well with our arbitrary taken magnitude of $\chi$ of SRDM), but most importantly for the validation tests, $\chi_{SnTe}$ shows a very strong temperature dependence. We benchmark $\chi_{SnTe}(T)$ first. To this end we measure a 58 mg large piece of the material, indicated in Figure 12 by the grey arrow. For this measurement, this large crystallite is fixed inside the S3 narrowed straw. One of its transverse dimensions sufficiently exceeds the internal diameter of this straw so the bulged straw provides the necessary grip to stabilize its location within the straw. The results are indicated as dark gray bullets in Figure 13.

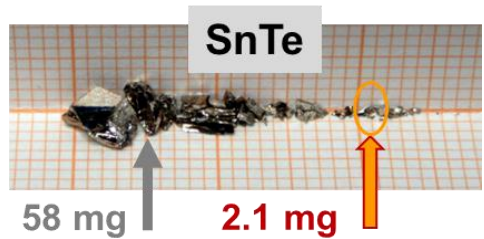

**Figure 12.** Selection of SnTe crystallites, leftovers from a several-cubic-centimeter large single crystal [45]. The arrows indicate two pieces used in this study: the larger for benchmarking of the temperature dependence of the magnetic susceptibility, the smaller for the validation of the *in situ* compensation method.

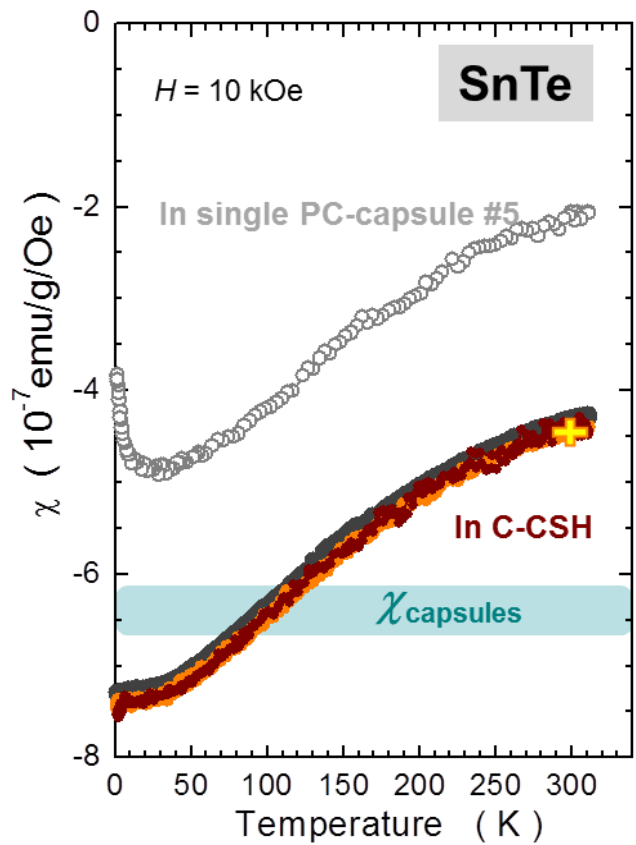

**Figure 13.** Temperature dependence of the magnetic susceptibility $\chi$ of SnTe established in this study by various methods. The background dark grey symbols mark the $\chi$-benchmarking measurement of a large 58 mg crystal. Open gray symbols

indicate the measurements of 2.1 mg crystallite in a single polycarbonate capsule #5. The orange symbols indicate the results obtained for the same minute crystallite mounted on a magnetically clean and uniform Si stick. The brown symbols indicate the results of the final method-validation test obtained when this small crystallite was mounted in the capsule compensated sample holder (C-CSH) elaborated in this study.

We next select the main test material, a small 2.1 mg SnTe crumb, indicated by the orange oval in Figure 12. First, we confirm that its magnetic susceptibility is identical to that of its larger neighbor. To accurately establish the absolute magnitudes of $\chi(T)$ of this tiny crystallite, we glue it to one of our standard Si sample holders. That is possible because it has two flat walls. We use an equivalently small droplet of strongly diluted GE-varnish [31]. We separately evaluate that at the same field of 10 kOe and in the whole $T$-range, the magnetic contribution of the solution used by us to fix the sample stays well below the error bars of the results obtained for the 2.1 mg sample, so it is disregarded in further data reduction. The results are marked by orange symbols in Figure 13, confirming the expected magnetic characteristics of the test piece.

The final two measurements are performed with this small test SnTe piece loaded (i) into a single PCC capsule, and (ii) into the central capsule of C-CSH. In both cases, we also measure the reference $T$-dependencies of the (i) empty capsule and (ii) empty C-CSH, which are subtracted accordingly. For both measurements the sample holder is centered on the bottom of the sample-housing capsule using the method outlined in Appendix A. All four measurements are done according to the same sequence. The results of the measurements performed in a single PCC are marked as open gray circles in Figure 13. They differ markedly from those two established previously. Obviously, the alignment error takes its toll here. We evaluated this error in Section 2.1.3 to be around 3 mg/mm for a standard diamagnetic material at the position corresponding to the bottom of the housing capsule (See Figure 4). Taking 0.5 mm as a reasonable estimation of the typical misalignment, one should expect about 1.5 mg uncertainty, which is about 70% of the mass of the test sample. Indeed, we find that the absolute values of $\chi_{SnTe}$ established in this approach differ by a factor of two at room temperature, which conforms favorably to our crude estimation. Certainly, with the identical alignment of the capsule for both these measurements. one should get a perfect reproduction of the dark grey (benchmark) or the orange results. In the authors' view, this is beyond the reach of the standard approach in the case of such minute signals. In contrast, the main part of the temperature dependence is captured quite reasonably, at least above approximately 50 K, where the $T$-responses of the $m$ of the capsules are reasonably flat. Below 50 K, where the PM contribution of capsules becomes substantial (c.f. Figure 2), a small difference in alignment prevents an adequate elimination of this signal and the specific magnetic properties of the capsule mar the true properties of the investigated material. It is worth noting that one can arrive at the last conclusion only after prior knowledge of the results.

A very different outcome is obtained when this tiny crystallite is investigated in the C-CSH. Importantly, as well as in the first case, the test sample is located at the bottom of a similar housing capsule, but the several-dozen times smaller sensitivity to misalignment makes the results obtained in C-CSH practically insensitive to this issue. The expected uncertainty band in this case is only 0.1 mg/mm, which for 0.5 mm inaccuracy yields the uncertainty as low as 2.5% of the mass of the test crystallite. This final step of the validation of the *in situ* compensation method is marked in Figure 13 using brown symbols. It must be noted that a precise one-to-one correspondence has been achieved, which strongly underlines the practical power of the *in situ* compensation and its potential. It shows that it can work very effectively in a broad range of environmental parameters, providing very accurate magnitudes of magnetic moments of even minute specimens of substances exhibiting very weak magnetic signatures.

One very important technical note is appropriate here, which highlights another practical advantage of the routine usage of the *in situ* compensation outlined in this report. We strongly emphasize that the compensation for the unbalanced signal of the empty C-CSH has been done by a simple subtraction of the results reported by the magnetometer in the respective *.dat files. The only restrictions to be enforced on the system are: (i) to per-form the *T*-dependence with the "Auto Tracking" option set on, and (ii) to use the "linear regression" algorithm. Apart from these, all the results reported here are obtained using the RSO method of measurements and the "Sweep" approach mode during the measurements in temperature domain. No Automatic Background Subtraction mode is necessary, neither is a cumbersome "scan-by-scan" subtraction of the reference measurements [10]. Any approach like the latter calls for excessive numerical effort and would likely cause delays in data reduction. In the method put forward here it is sufficient to operate only on the basic output *.dat files yielded by the magnetometer. In the authors' view, this very convenient property is the result of the perfectly even symmetry of the SRF and the unprecedented insensitivity to small alignment errors. The first fitting constricts channels least mean square (LMS) routines, in their fitting effort, onto specific even-parity components in the signal under processing. That is why it is impossible to correctly extract the magnitude of a weak moment located at the bottom of a single capsule, as considered in Section 2.1.3. Colloquially speaking, there is still too much of even-parity signal of the capsule itself at its bottom that this contribution takes over the searched weak source of signal. In contrast, as shown in Figure 8c, the response of the same capsule abutted by other similar capsules is highly complicated, without any dominant parity. Therefore, the same weak signal effectively breaks out and is quite correctly evaluated by LMS procedures, as exemplified in the following example.

In order to illustrate why the approach described above yields so surprisingly accurate magnitudes of very weak magnetic moments, we scrutinize the exact form of the scans recorded during these two measurements: the *T*-dependencies of the empty C-CSH and the same C-CSH with the 2.1 mg test piece of SnTe loaded to the central capsule. As the sole representatives, we take the very first points of these two measurements. For both measurements the C-CSH assembly was centered once, i.e., before the entire validation test, using the method described in Appendix A. There is generally no need for a second centering (correction) after inserting (removing) the sample into (from) the central capsule and reassembling the C-CSH. Just the basic care exerted during these processes allows the reconstruction of the identical enough C-CSH; a repeated centering is not needed. Moreover, the centering process on its own does not provide more accuracy than the careful reassembly of the C-CSH. Our experience gained along the development process, substantiated by the analysis of the misalignment errors (Figure 9), says that 0.5 mm reproducibility is probably the best possible result and that it is sufficient.

In Figure 14a, open light grey circles and dark grey bullets mark the shape of the response of the empty C-CSH and with 2.1 mg test crystallite loaded to the central capsule, respectively. Clearly, none of the scans resemble the typical shape of the SRF. The orange bullets represent the calculated point-by-point difference between these two scans yielding the sole contribution brought about by the insertion of the test sample. This shape exhibits a very clear correspondence to the signal expected for a diamagnetic substance, and the solid red line marks the fit of the SRF centered at the center of the scan window, i.e., where the whole assembly was centered. The match is exceptionally good and the *z*-shift of about 1 mm indicates the inaccuracy in alignments of the C-CSH assembly for these two measurements. The fit yields $m_{SnTe}$ = −9.64 μemu. The corresponding magnitude of the magnetic susceptibility is marked by a yellow cross in Figure 13. This result corresponds to the ABS methods at its best.

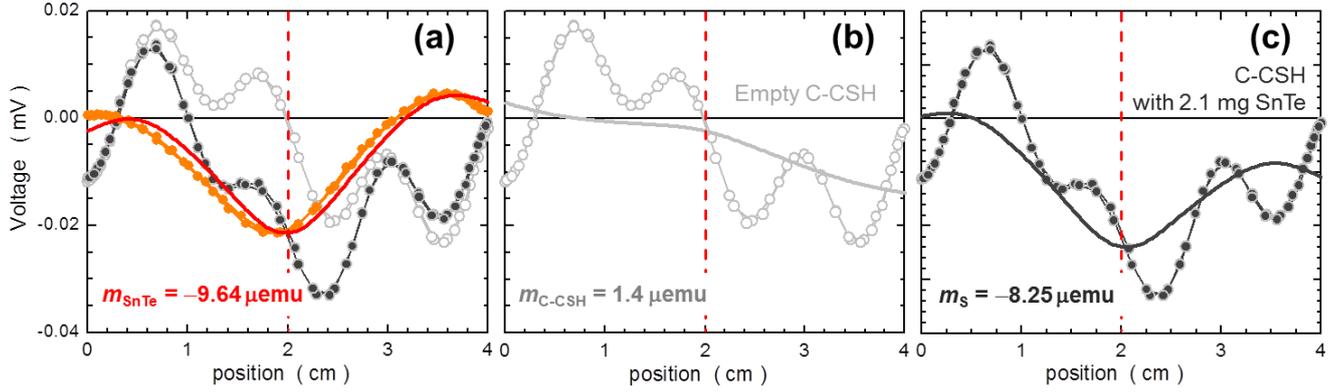

**Figure 14.** Exemplary scans (the long averaged scaled response voltages) representing the first experimental point of the temperature dependence of magnetic susceptibility of a 2.1 mg SnTe crystallite depicted by brown symbols in Figure 13. Light grey open symbols in (**a**) and (**b**) mark the scan of the empty capsule compensational sample holder (C-CSH). Dark grey symbols mark the scan obtained with the 2.1 mg SnTe piece installed in the central capsule of the C-CSH. The orange symbols in (**a**) mark the difference between scans presented in (**c**) and (**b**). The solid lines in matching colors in all three panels represent least mean square fit of the SQUID response function (defined in Appendix B) to the experimental points. The corresponding magnetic moments $m$ are given in the labels.

Now we separately extract the magnetic moments from each of the two experimental scans given in Figure 14a. They are plotted in Figures 14b,c, together with the corresponding fits, marked as solid lines of matching colors. In the first case, Figure 14b, the fit yields the residual moment of the empty C-CSH $m_{C-CSH}$ = +1.4 μemu. In the second case, Figure 14c, after insertion of the sample the fit yields $m_S$ = −8.25 μemu. Obviously, at first glance, neither of the two fits can be classified as satisfactory. The fitted functions cut somewhat freely across the experimental points, taking rather little notice of the pattern formed by the experimental points. This is actually acceptable, as the fitting routine seeks the even-parity component, the SRF, and only these "frequencies" that correspond to $z_0$ = 2 cm. It is therefore no surprise that the difference of the magnitudes of $m$ returned by the fittings yields the correct magnitude of $m_{SnTe} = m_S − m_{C-CSH}$ = −9.65 μemu, a mere 0.1% different value than established in Figure 14a.

Finally, we exemplify the usefulness of the in situ compensation method in biological and medical studies by performing measurements of common turmeric powder. We emphasize that the investigated sample has not been prepared in any standard manner, e.g., Association of Analytical Chemists standard methods, specific for biological substances. The powder originates from a commercial herbs retailer [47]. A quantity of 2.9 mg of the powder is loaded to the #5 PCC and measured in C-CSH. This amount of material forms about a 1.5 mm thin layer occupying the bottom of the capsule; however, some of the powder spreads all over the capsule. As is shown in Figure 3, a nearly fully loaded #5 PCC accommodates approximately 70–80 mg of fresh turmeric powder. We do not focus on absolute magnitudes of $m$ in this measurement., primarily because after loading to the SQUID sample chamber, a weak drift in the magnetic signal is observed. Similar to the gelatin, water evaporates form the fresh turmeric powder in the dry environment of the SQUID sample chamber, so for fully quantitative studies the proper preparation of such a material is mandatory. In this measurement, we aim at properties at the low temperatures end. All the data presented below are corrected for the residual weak signal of the empty C-CSH.

We start from room temperature $m(H)$ presented as orange symbols in Figure 15b. The negative slope confirms the diamagnetic character of turmeric, although a small near zero feature indicated that a small amount of ferromagnetic (FM) component is present. This may well be a contamination due to preparation of the spice at its place of origin. From the slope of this $m(H)$, we evaluate the diamagnetic susceptibility as $\chi_{turmeric} \cong -3.3 \times 10^{-7}$ emu/g/Oe. The results of the $T$-dependence of $m$ presented in Figure 15a indicate that turmeric turns from a diamagnetic substance into the paramagnetic (PM) one at low temperatures. To quantify this PM component, we measure $m(H)$ at two very low temperatures: 1.8 and 5 K, shown in Figure 15b. Clearly, the two dependencies have a Brillouin-like character at high magnetic fields, with noticeable FM-like deviations around $H = 0$, similar to that at 300 K.

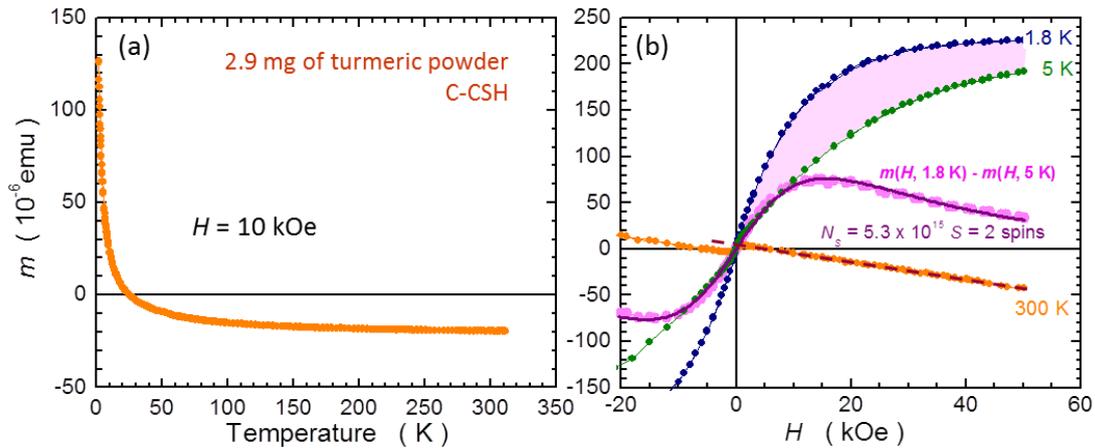

**Figure 15.** Results of the magnetic studies of 2.9 mg of common turmeric powder performed in capsule compensated sample holder (C-CSH). (**a**) Temperature dependence of magnetic moment $m$ measured at external magnetic field $H = 10$ kOe. (**b**) Values of $m(H)$ at 300, 5, and 1.8 K. The brown dashed line yields the magnitude of diamagnetic susceptibility of the sample $\chi_{turmeric} \cong -3.3 \times 10^{-7}$ emu/g/Oe. The magenta area and bullets mark the difference between $m(H)$ at 2 and 5 K. The solid purple line indicates difference between Brillouin functions calculated for $S = 2$ (Equation (A2) in Appendix D). The matching factor between the theory and experiment, $N_S = 5.3(1) \times 10^{15}$, yields the number of paramagnetic $S = 2$ spins in the sample.

In order to accurately separate the PM signal in our turmeric sample, we invoke the fact that in most cases it is only the PM component that is responsible for changes of $m(H)$ at low temperatures, a condition met in many magnetically composite systems in which the technique presented below worked with a very high precision [3,48–50]. The diamagnetism should not depend on $T$, so should most of the FM contaminations, as latter are typically characterized by a very high spin coupling temperatures. Indeed, the $m(T)$ dependence does not indicate any magnetic phase transition below 300 K. In such circumstances the calculated difference between the experimental points obtained at these two low temperatures should only reflect the change in the PM component—all other contributions get canceled. The magenta filled area and the corresponding symbols in Figure 15b represent this difference. The required information on the spin state is obtained from fitting of this difference by the theoretical formulae describing the difference of the two Brillouin functions taken at 1.8 and 5 K. Its particular form and relevant technical details are given in Appendix D. The best fit is obtained for $N_S = 5.3(1) \times 10^{15}$ of $S = 2$ spins in the sample, which indicates their mass concentration $x_S \cong 1.8 \times 10^{19}$ per gram of this particular powder. The other spin values of $S = 3/2$ and $5/2$ definitely yielded worse fits. The value of $S = 2$ indicates $Fe^{4+}$ ions, as this transition metal dominates by far the other transition metals in turmeric [51]. This analysis quantifies the dominant contribution; it does not preclude the presence of other unpaired spins in the material. Commercial magnetometers are volume probes and are not element specific, so they are limited in this respect. However, although this result cannot be regarded as a standard for the turmeric powder, it is only the employment of the C-CSH that makes such as analysis quantitative. It has to be noted that, as shown in Figure 2, the capsules themselves do exhibit PM properties, and this PM of capsules would mar the results of the analysis presented above. In contrast, the self-compensation of the magnetic properties of capsules in the C-CSH assembly eradicates this detrimental property, as documented for both kinds of capsules in Figure 10 in the $T$-domain and in Figure 11 in the $H$-domain.

Where the magnetic fields are concerned, it has to be noted that the "stealthy" properties of C-CSH also lead to much smaller errors of $m$ caused by the trapped field of the superconducting magnet [19,20,52–54]. They are also reduced 20 to 30 times, from a level of about $+4 \times 10^{-7}$ emu when the specimen is housed in a single capsule to the magnetometer noise level of 10$^{-8}$ emu in C-CSH. Note that the shift of *m* due to the presence of uncompensated capsules on the return from a trip to high positive fields causes a shift of *m* towards positive values and it will be negative on the trip back from high negative fields, adding a FM-like hysteresis feature to the results of considerably high height of about 8 × 10$^{-7}$ emu, an important value to address in the precise magnetometry of weak signals.

## 5. Conclusions

The construction principle, assembly method, validation measurement, and example of employment of the *in situ* compensation concept applied to commonly used capsules in investigations of small and irregular shape fragments, powders, and other specimens that would generally require complicated mounting in SQUID magnetometers have been presented. It has been experimentally evidenced that with incorporation of only the basic components, familiar and ubiquitously present in every magnetometry lab—common plastic straws and capsules (most preferably the polycarbonate ones)—an assembly can be completed in virtually one afternoon, which allows one to reduce the magnetic signatures of a specimen-housing capsule at least 20 times in time, temperature, and magnetic field domains. In practical terms, the reduction is equivalent to the reduction of the mass of the capsule to no more than 1.5 mg, sometimes even twice that. This permits investigation of larger amounts of materials, say between 50 and 150 mg, without any need for background signal subtraction—a procedure that is particularly prone to a large experimental inaccuracy caused by a non-ideal alignment of the sample holder. However, the effort exerted in preparation of the capsule compensated sample holder is particularly useful when only minute amounts of material are available for magnetometry. Such specimens occupy only the bottom part of the housing capsule and the background subtraction error due to misalignment can be comparable to the mass of the specimen, making the magnetometry semi-qualitative, at best. In C-CSH, the magnetic signal at the location of the bottom of the housing capsule can be readily reduced up to 30 times, with alignment errors not exceeding 0.1 mg of an equivalent mass of a common diamagnetic material. This opens a new experimental window of truly precision magnetometry of miniscule specimens of masses ranging down to a single milligram. This unprecedented property has been validated by demonstration of the ability to obtain accurate absolute values of the temperature dependence of the magnetic susceptibility of a tiny, a mere 2.1 mg, crystallite of diamagnetic SnTe.

A full applicability of the C-CSH concept to small amounts of powdered samples has been presented by investigation of a similarly minute sample of common turmeric powder. Both the magnitude of the diamagnetism of the substance and of the paramagnetic contribution due to iron ions were readily obtained from magnetic field dependencies measured at room temperature and at liquid helium temperatures, respectively. All the measurements were obtained in one experimental session, without remounting the specimen to different (specialized) sample holders. Our effort proves that the C-CSH concept works with high reproducibility rate between 1.8 and 320 K and at any magnetic field available in the magnetometer.

The concept of the *in situ* compensation of capsule containers has been tested in MPMS-XL SQUID magnetometers of Quantum Design. The authors are confident that after adequate modifications, a similar assembly may prove equally effective in other types of magnetometers, although for the VSM-type MPMS 3 system one would like to see even smaller polycarbonate capsules that the smallest ones of size 5 available today.

## 6. Patents

Relevant patent application has been registered with Polish Patent Office as P.439874.

**Funding:** This study has been supported by the National Science Centre (Poland) through project OPUS (DEC–2017/27/B/ST3/02470).

**Acknowledgments:** The authors acknowledge gratefully Andrzej Szczerbakow for providing SnTe samples, Piotr Nowicki for technical support and Jacek Szczytko and Marc Kunzmann for providing us the capsules needed for this study.

**Appendix A. (Centering of C-CSH)**

Setting up the position of the sample in the sample chamber—the alignment—is yet another prerequisite for the precise magnetometry. In the case of capsules housing miniscule specimens, we cannot rely on the bare voltage related to the moment of the sample as it is dwarfed by that of the capsule. Obviously, an additional magnetic marker has to be used that can be affixed temporarily to the sample holder (outside wall of the straw in this case). The authors do not recommend using pieces of magnetically strong ferrous materials as the positioning could only be done at very weak magnetic field, which calls for unnecessary frequent changing of the field in the magnetometer and related field instabilities. Our material of choice is a small, about 10 mg, crystallite of PbTe weakly doped with Cr. In a semiconductor matrix, chromium tends to precipitate to form ferromagnetic nanocrystals [55], which, as an ensemble, exhibit superparamagnetic properties even at room temperature [49], yielding a convenient, approximately 100 times stronger, response at $H$ = 10 kOe than anything else considered in this study. An adequate amount of any similar composite material can be used for this purpose.

A narrow 2–3 cm long strip of a commercial surgical tape with acrylic adhesive serves as the bonding agent for the marker. The tape sticks firmly to the straw and comes off easily without leaving any residue. A freshly cut tape exhibits a weak magnetic response, works well down to helium temperatures, and can be considered as a very convenient "fist-aid" fixing medium in magnetometry for medium to strong magnetic materials. An example of our method is given in Figure A1.

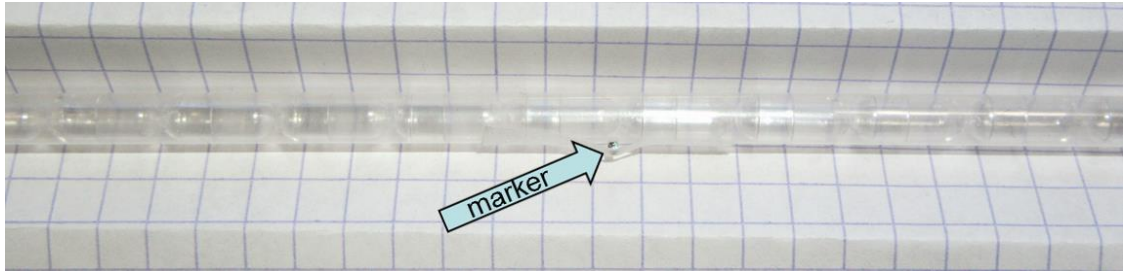

**Figure A1.** An example of the application of a magnetic marker for centering of the capsule compensational sample holder. As indicated, it is hard to get better reproducibility than about 0.5 mm.

**Appendix B. External Fitting of the SQUID Scan Voltage**

The necessary position dependence of the SQUID output during scanning can be extracted from the relevant *.raw file(s), or can be taken from the *.lastscan file, both provided by the MultiVu software. For any studies focussed on precise magnetometry, we strongly advise recording the full suite of the data, that being the accompanying *.diag and *. raw files. On numerous occasions, they proved indispensable in troubleshooting or error correction. Alternatively, the required scans can be measured purposely using, for example, the [Measure] tab of MultiVu program. Among many available software packages, we use Origin's NonLinear Curve Fitting (NLCF) option, and to straightforwardly extract the magnitude of magnetic moment *m* we use the "Long Avg. Scaled Response", that is, the 17th column in the *.lastscan file. Our fitting function $V(z)$ comprise the second-order axial gradiometer response function (SRF with a linear term added to compensate for the offset and even-parity signal drifts) [56]:

$$V(z) = mr^3 \left\{ \frac{2}{(r^2 + (z-z_0)^2)^{\frac{3}{2}}} - \frac{1}{[r^2 + (z-z_0+a)^2]^{\frac{3}{2}}} - \frac{1}{[r^2 + (z-z_0-a)^2]^{\frac{3}{2}}} \right\} + A + Bz, \qquad (A1)$$

where $r$ = 0.97 cm is the radius of the gradiometer pick up coils, $a$ = 1.519 cm is their base distance, and $z_0$ is the axial position of the sample (the alignment). In the coordinates used by MultiVu, $z_0$ should be close to 2 cm. The form given in Equation (A1) means that the numerical analysis is done in point dipole approximation. Therefore, for extended specimens the final magnitudes of *m* are corrected for the shape-dependent factors [19,43,44,57]. The main advantage of the external analysis of the scans is that the user possesses the ability to freely adjust with the parameters of *V(z)*. Most importantly here, one can investigate the dependence of *m* on the alignment.

**Appendix C. RSO Method Settings**

As noted in the Materials and Methods section, in most cases the data presented here are collected by taking six measurements each consisting of four oscillations (24 scans all together to be averaged). There are two practical hints concerning these parameters that are worth adding to this report. First, for a given number of single scan events (4 × 6 = 24 in this case), it does not make any difference if all of them constitute one single measurement unit or, conversely, whether they are divided into 24 separate measurements containing only one scan event. The average result will be the same. However, for advanced users, particularly those who are using post-measurement method(s) to reanalyze the scans, it pays off to have more measurement units than oscillation events and to record the whole suite of the output files provided by MultiVu software (.dat, .diag, and .raw files). The advantage becomes evident when, for example, say during the ninth oscillation event, a sudden signal jump on the scan voltage occurs. This distorted voltage is "normally" averaged together with the other, correct readings (scans)—resulting in a false moment determination, accompanied by an increased magnitude of the standard deviation error. However, for the case considered here, this jump mars only the third measurement unit. A short inspection of the corresponding .diag file, manual or by a dedicated software, can eradicate such wrong readings and after recalculation of the new average, the true (or more correct) magnitude of *m* at the given experimental conditions is restored. This will be accompanied by a reduction of the standard deviation error down to the level of the neighboring points.

**Appendix D. Extraction of a Number of PM Spins**

Accurate quantitative information on the paramagnetic contribution present in a composite material can be conveniently obtained from isotherms *m(H)* measured at very low temperatures ($T_1$, $T_2$, here $T_1$ = 1.8 K, $T_2$ = 5 K) and numerical analysis of their difference Δ*m(H, ΔT = $T_1$ − $T_2$)*. The values of $T_1$ and $T_2$ should be much smaller than the spin–spin coupling temperature(s) of the remaining non-PM part of the sample. This allows one to treat the diamagnetic, ferromagnetic, and all other magnetic contributions as temperature independent, and so, whatever is the form of their contribution they null out in Δ*m(H, ΔT)*. Under such circumstances, the analysis does not have to account for them and the difference Δ*m(H, ΔT)* (marked by magenta symbols in Figure 15b) can be fitted by:

$$\Delta m(H, \Delta T) = N_S g \mu_B S \left[ B_J(H, 1.8\text{K}) - B_J(H, 5\text{K}) \right] \tag{A2}$$

where $B_J(H, T)$ is the Brillouin function:

$$B_J(x) = \frac{2J+1}{2J} \coth\left(\frac{2J+1}{2J}x\right) - \frac{1}{2J} \coth\left(\frac{1}{2J}x\right)$$

with *H* and *T* tied in:

$$x = J \frac{g\mu_B H}{k_B T} \tag{A3}$$

Here, g = 2 is the Land'e factor, $\mu_B$ is the Bohr magneton, $k_B$ is the Boltzmann constant, and $N_S$ is the number of spins with *J = L + S*. For simplicity we take *L* = 0. In this approach the magnitudes of *S* and $N_S$ are the only adjustable parameters. As *S* can only assume some discrete values, in a search of the spin state we perform three independent fittings for *J = S* = 3/2, 2, and 5/2. The best fit, obtained for *S* = 2 and $N_S$ = 5.3(1) × 10$^{15}$, is presented in Figure 15b.

## 7. Glossary

List of nonstandard terms and abbreviations used in this paper.

| Abbreviation | Meaning |
| --- | --- |
| ABS | automatic background subtraction |
| C-CSH | capsule compensational sample holder |
| GC | gelatin capsule |
| LMS | least mean square |
| PCC | polycarbonate capsule |
| QD | Quantum Design |
| RSO | reciprocal space option |
| SRDM | standard reference diamagnetic material |
| SRF | SQUID response function |